\documentclass[letter]{ieice}
\usepackage[dvips]{graphicx}

\field{}
\title{Towards Interactive Object-Oriented Programming}
\authorlist{
\authorentry{Keehang KWON}{n}{labelA}
\authorentry{Kyunghwan Park}{n}{labelB}
\authorentry{Mi-Young Park}{m}{labelA}
}
\affiliate[labelA]{The authors are with Computer Eng., DongA Univ., Korea.
\{khkwon,openmp\}@dau.ac.kr}
\affiliate[labelB]{The author is with Computer Eng., DongA Univ. khpark@dau.ac.kr.
The corresponding author.}

\newenvironment{numberedlist}
{\begin{list}{\makebox[20pt]{\hss(\arabic{itemno})\enspace}}
             {\usecounter{itemno}\labelwidth 20pt}}{\end{list}}

\newcounter{itemno}

\newcounter{itemno1}

\newcounter{itemno2}

\newcounter{exno}

\newcounter{defno}

\newenvironment{defn}{\refstepcounter{defno}\medskip \noindent {\bf
Definition \thedefno.\ }}{\medskip}

\newcommand{\sep}{\;\vert\;}

\newcommand{\oprove}{\vdash\kern-.6em\lower.7ex\hbox{$\scriptstyle O$}\,}

\newcommand{\Pscr}{{\cal P}}

\newcommand{\all}{\forall}

\newsavebox{\lpartfig}
\newsavebox{\rpartfig}

\newenvironment{exmple}{
 \begingroup \begin{tabbing} \hspace{2em}\= \hspace{3em}\= \hspace{3em}\=
\hspace{3em}\= \hspace{3em}\= \hspace{3em}\= \kill}{
 \end{tabbing}\endgroup}

\newcommand{\lb}{\langle}
\newcommand{\rb}{\rangle}

\newcommand{\add}{\sqcup} 

     {\\* \hspace*{\fill} \end{trivlist}}

\begin{document}
\maketitle
\begin{summary}
To represent interactive objects,  we propose
 a choice-disjunctive declaration statement of the form
$S \add R$ where $S, R$ are the (procedure or field) declaration statements within
 a class.
This statement has the
following semantics: request the user to choose one between $S$ and $R$ when an
object of this class is created.
  This statement is useful for representing interactive objects that require
interactions with the user.
\end{summary}
\begin{keywords}
interactions, object-oriented, computability logic.
\end{keywords}

\section{Introduction}\label{sec:intro}

Interactive  programming \cite{Lyn96,Rol10} is an important modern trend in information technology.
 Despite much popularity, object-oriented languages \cite{Avi03,Jos12,Jos08} have
traditionally lacked mechanisms  for representing interactive objects.
For example, an object like a lottery ticket is in a superposition state of
several possible values and require further interactions with the environment to determine
their final value.

To represent interactive objects,  we propose to adopt a choice-disjunctive
operator in computability logic \cite{Jap03,Jap08}. To be specific,
we allow, within a class definition,
 a choice-disjunctive declaration statement of the form
$S \add R$.
This statement has the
following semantics: request the user to choose one between  $S$ and $R$ when
an object is created.
  This statement  is useful for representing interactive objects. For example,
a lottery ticket, declared as $value = \$0 \add\ value = \$1M$, indicates that it
has two possible values, nothing or one million dollars, and its final value will
be determined by the environment (or the user).

The remainder of this paper is structured as follows. We describe the new language $Java^i$
 in
the next section. In Section \ref{sec:modules}, we
present some examples.
Section~\ref{sec:conc} concludes the paper.

\section{The Language}\label{sec:logic}

The language is a subset of the core (untyped) Java
 with some extensions. It is described
by $G$- and $D$-formulas given by the syntax rules below:
\begin{exmple}
\>$G ::=$ \>   $A \sep x = E \sep  G;G \sep o = new\ D$ \\   \\
\>$D ::=$ \>  $ A := G  \sep x = E \sep \all x\ D \sep D \land D \sep D \add D$\\

\end{exmple}
\noindent
In the rules above,  $o$ is an object name, $x$ is a field name, $E$ is an expression, and
$A$  represents a procedure (or a method) of the form $p(t_1,\ldots,t_n)$.
The notation $x = E$ in $G$ denotes an assignment statement.

A $D$-formula  is called a class definition. The notation
$x = E$ in $D$ denotes a field $x$ with an initial
value $E$. The notation
$A := G$ in $D$ denotes a procedure declaration where $G$ is called a procedure body.
The notation $D\land D$ denotes a conjunction of two $D$-formulas.

In the transition system to be considered, $G$-formulas will function as the
main program (or procedure bodies), and a set of tuples $\lb o,D\rb$ where $o$ is an object
name and $D$ is a $D$-formula will constitute  a program.

 We will  present an operational
semantics for this language via a proof theory. The rules  are formalized by means of what
it means to
execute the main task $G$ from a program $\Pscr$.
These rules in fact depend on the top-level
constructor in the expression,  a property known as
uniform provability\cite{MNPS91}. Below the notation $\lb o,D\rb;\Pscr$ denotes
$\{ \lb o,D\rb \} \cup \Pscr$ but with the $\lb o,D\rb$ tuple being distinguished
(marked for backchaining). Note that execution  alternates between
two phases: the goal-reduction phase (one  without a distinguished tuple)
and the backchaining phase (one with a distinguished tuple).
The notation $S\ sand\ R$ denotes the following: execute $S$ and execute
$R$ sequentially. It is considered a success if both executions succeed.
The notation  $o.G$ represent an association of $o$ with every field or procedure
name appearing in $G$.
For example, if $G$ is $p(t_1,\ldots,t_n)$, then $o.G$  represents  $o.p (t_1,\ldots,t_n)$.

\begin{defn}\label{def:semantics}
Let $o$ be an object name, let $G$ be a main task and let $\Pscr$ be a program.
Then the notion of   executing $\lb \Pscr,o.G\rb$ successfully and producing a new
program $\Pscr'$-- $ex(\Pscr,o.G,\Pscr')$ --
 is defined as follows:
\begin{numberedlist}

\item    $ex(\lb o,(A := G)\rb;\Pscr,A,\Pscr')$ if
 $ex(\Pscr, o.G,\Pscr')$ . \% matching procedure for $A$ is found

\item    $ex(\lb o,\all x D\rb;\Pscr,A,\Pscr')$ if   $ex(\lb o,[t/x]D \rb;
\Pscr, A,\Pscr')$. \% argument passing

 \item    $ex(\lb o, D_1\land D_2\rb;\Pscr,A,\Pscr')$ if   $ex(\lb o,D_1\rb;
\Pscr, A,\Pscr')$. \% looking for the procedure $A$ in $D_1$.

\item    $ex(\lb o, D_1\land D_2\rb;\Pscr,A,\Pscr')$ if   $ex(\lb o,D_2\rb;
\Pscr, A,\Pscr')$. \% looking for  the procedure $A$ in $D_2$

\item    $ex(\Pscr,o.A,\Pscr')$ if   $\lb o,D\rb \in \Pscr$ and
$ex(\lb o,D\rb;\Pscr, A,\Pscr')$. \% a procedure call in object $o$

\item  $ex(\Pscr,o.x = E,
\Pscr')$ where
$\Pscr'$ is obtained from $\Pscr$ by first evaluating $E$ to $E'$ and updating
the value of the field $x$ to $E'$ in the object $o$.

\item  $ex(\Pscr,G_1; G_2,\Pscr_2)$  if $ex(\Pscr,G_1,\Pscr_1)$  sand \\
  $ex(\Pscr_1,G_2,\Pscr_2)$.

\item  $ex(\Pscr,o = new\  D, \{ \lb o,D' \rb \} \cup\Pscr)$ where $D'$ is obtained from $D$ by
first removing choice-disjunctions and then by initializing its fields. \% object creation

\end{numberedlist}
\end{defn}

\noindent
If $ex(\Pscr,G,\Pscr_1)$ has no derivation, then the machine returns the failure.
In the above, the rules (1) to (4) deal with the backchaining phase, whereas the rules (5) to (8) deal
with the goal reduction phase.
Our operational semantics is a standard one appearing in most textbooks.
Only the rule (8) is a novel feature.

\section{Examples}\label{sec:modules}

Imagine Temple University charges \$5,000 as its tuition for nonemployees
and \$3,000 for employees.
An example of this class is provided by the
following program:

\begin{exmple}
 $class\ TempleU$\\
$tuition = 0\ \land$ \\
$(employee = true \add employee = false)\ \land$\\
$(comp\_tuition() :=  if\ employee\, then\ tuition = \$3000\ $\\
$ \hspace{10em} else\ tuition = \$5000) $\\
$void\ main()$\\
$TempleU\ p = new\ TempleU$; \\
$comp\_tuition()$;\\
$print(p.tuition)$
\end{exmple}
\noindent In the above, creating a TempleU object via the $new$ construct
 basically proceeds as follows: the machine asks the user ``are you an employee?''.
If the user answers yes by choosing the left disjunct, $employee$ will be initialized to
$true$ and the machine will eventually print \$3000 for its tuition.
 If the user answers no by choosing the right disjunct,  $employee$  will be initialized
 to $false$ and the machine will eventually print \$5000 for its tuition.
Our language thus makes it possible to customize the amount for tuition via
interaction with the user.

\section{Conclusion}\label{sec:conc}

In this paper, we have considered an extension to the core Java with
disjunctive statements within a class definition. This extension allows statements of
the form  $S \add R$  where $S, R$ are statements.
These statements are
 particularly useful for representing interactive objects.

\section{Acknowledgements}

This work  was supported by Dong-A University Research Fund.

\bibliographystyle{ieicetr}


\end{document}